Tunnel magnetoresistance and robust room temperature exchange bias

with multiferroic BiFeO$_3$ epitaxial thin films

H. Béa

Unité Mixte de Physique CNRS-Thales, Route départementale 128, 91767 Palaiseau, France

M. Bibes

Institut d'Electronique Fondamentale, CNRS, Université Paris-Sud, 91405 Orsay, France

S. Cherifi

Laboratoire Louis Néel, CNRS, BP166, 38042 Grenoble, France

F. Nolting

Paul Scherrer Institut, 5232 Villigen-PSI, Switzerland

B. Warot-Fonrose

CEMES, CNRS, 12 rue Jeanne Marvig, 31400 Toulouse, France

S. Fusil, G. Herranz, C. Deranlot, E. Jacquet, K. Bouzehouane and A. Barthélémy

Unité Mixte de Physique CNRS-Thales, Route départementale 128, 91767 Palaiseau, France

Abstract

We report on the functionalization of multiferroic BiFeO$_3$ epitaxial films for spintronics. A first example is provided by the use of ultrathin layers of BiFeO$_3$ as tunnel barriers in magnetic tunnel junctions with La$_{2/3}$Sr$_{1/3}$MnO$_3$ and Co electrodes. In such structures, a positive tunnel magnetoresistance up to 30% is obtained at low temperature. A second example is the exploitation of the antiferromagnetic spin structure of a BiFeO$_3$ film to induce a sizeable (~60 Oe) exchange bias on a ferromagnetic film of CoFeB, at room temperature. Remarkably, the exchange bias effect is robust upon magnetic field cycling, with no indications of training.

Multiferroic materials [1,2], that display several long-range orders among ferroelectricity (FE), ferromagnetism (FM) and ferroelasticity, can bring interesting additional functionalities to spintronics [3]. For instance, the magnetoelectric coupling existing in these compounds could be used to rotate the magnetization of a ferromagnetic multiferroic element by applying an electric field, rather than a magnetic field. Unfortunately, very few multiferroics exhibit a finite magnetization (i.e. are ferromagnetic or ferrimagnetic). Most of them are antiferromagnetic (AF) or weak-ferromagnets. Still, many spintronics applications can also be devised exploiting the new functionalities provided by antiferromagnetic multiferroics (AFM). For example, Binek and Doudin [4] have proposed to use a thin AFM layer as a tunnel barrier in a FM/AFM/FM magnetic tunnel junction (MTJ). In this type of device, the magnetoelectric coupling is used to tune the effective direction of the exchange bias by an electric field. In other words, the magnetic configuration of the MTJ can be changed from parallel (P) to antiparallel (AP) (hence the resistance level from low to high) by applying an electric field across it. This type of device looks very appealing for spintronics since it would allow to write a magnetic bit electrically with a low power consumption, which is a major challenge for the future of magnetic random access memories (MRAMs) and other spin-based devices.

Following the suggestions of Binek and Doudin, we report here on the use of multiferroic $BiFeO_3$ (BFO) epitaxial films as tunnel barriers in MTJs and as exchange-biasing layers. We show that these BFO films have an excellent structural quality when integrated into perovskite-based spintronics architectures and display an AF behavior combined with ferroelectric properties. A positive tunnel magnetoresistance (TMR) of up to 30% is obtained at low temperature in $CoO/Co/BFO/La_{2/3}Sr_{1/3}MnO_3$(LSMO) MTJs. Finally, we show that such BFO layers can also be used to induce a robust exchange-bias of ~60 Oe at room temperature on $Co_{72}Fe_8B_{20}$ (CoFeB), a high Curie temperature ($T_C$) highly spin-polarized soft ferromagnet, widely used in last generation MTJs using MgO tunnel barriers [5].

We have grown BFO thin films and BFO/LSMO(15nm) heterostructures on (001)-oriented $SrTiO_3$ (STO) by pulsed laser deposition [6,7]. To define a MTJ, we then sputtered on a BFO(5nm)/LSMO(15nm) bilayer a top electrode of Au/CoO/Co and etched 30x30µm² junctions [8]. Figure 1 displays a high resolution transmission electron microscopy (TEM) micrograph of a BFO(8nm)/LSMO(15nm)//STO sample observed in cross-section on a Tecnai F20 microscope equipped with a spherical aberration corrector. An epitaxial growth of the LSMO layer on the STO with the usual epitaxial relations is observed and no dislocations are detected in the BFO layer. This has been checked in different parts of the sample. The crystalline quality of the interface, which is a key parameter for tunnel-type transport [9], is excellent, even though it is impossible to rule out the presence of some oxygen vacancies. The electrical quality of the barrier has been checked previously by conducting–tip atomic force microscopy [7].

We have measured the resistance R of several CoO/Co/BFO/LSMO junctions as a function of a magnetic field H applied along the [100] direction after field cooling (H=6 kOe) along the same direction. R(H) curves measured at 3K and a bias of 10 mV show a clear positive TMR for several junctions with 2 or 5 nm thick BFO barriers, see an example for a TMR of about 14% (calculated form TMR=($R_{ap}$-$R_p$)/$R_p$ with $R_{ap}$ and $R_p$ the antiparallel and parallel resistances respectively) in figure 2a. The resistance switchings at H = –170 Oe and 165 Oe correspond to the reversal of the LSMO magnetization and the ones at 350 Oe and – 1110 Oe to the reversal of Co exchange-biased by CoO. This positive TMR of 14% reflects a positive spin-polarization of $P_{Co}$=7.3% for Co at the interface with BFO (from Jullière formula [10] and using a spin polarization P of 90% for LSMO [8,12]). The maximum TMR we have obtained in such junctions amounts to +31.3% ± 2.4%, yielding a maximum $P_{Co}$=15.0% ± 1.0%. Previous TMR studies on Co/SrTiO$_3$/LSMO [11], Co/LaAlO$_3$/LSMO [12], Co/TiO$_2$/LSMO [13] and Co/Al$_2$O$_3$/LSMO [11] have shown that the sign and amplitude of the TMR depends on the wavefunction matching at electrode-barrier interfaces and on the electronic structure of the barrier. The positive TMR we observe here could thus reflect the preferential transition of positively spin-polarized Co wave functions by the epitaxial BFO barrier (the complex electronic structure [14,15] of which has not been calculated yet). Alternatively, it is possible that symmetry mixing occurs at the interfaces due to the presence of structural imperfections such as oxygen vacancies. Further experimental work and theoretical input on this point are required for a deeper understanding.

A study of the TMR as a function of the temperature shows that it decreases with temperature and vanishes around T*=200K (fig. 2b), despite the rather high $T_C$ of the LSMO electrode (330K) deduced from the magnetization versus temperature curves. This discrepancy between T* and $T_C$ in manganite tunnel junctions is a long-standing issue [16,17]. It was shown previously that the $T_C$ of high quality manganite interfaces is around 300K [18,19], i.e. lower than the bulk $T_C$ by some 50K. The lower value of T* that we observe here can also signal a local deoxygenation of the LSMO surface during the low-pressure growth of the BFO layer (6.10$^{-3}$ mbar vs 0.41 mbar for LSMO).

We conclude this part on BFO-based tunnel junctions by noting that the maximum TMR value we have obtained is rather large (comparable in absolute value to those measured in Co/LaAlO$_3$/LSMO [12] or Co/STO/LSMO junctions [11]), despite the AF nature of the tunnel barrier (see later). Previous measurements on MTJs with AF manganite barriers (e.g. La$_{0.55}$Ca$_{0.45}$MnO$_3$) had yielded rather low TMR and spin-polarization values [20,21]. This can be due to either spin depolarization at interfaces (likely to present a large degree of spin disorder due to a competition between FM interactions – in the electrode – and AF interactions – in the barrier) or to spin depolarization during the tunneling process (e.g. via magnons excitation). Even though systematic studies on junctions with different barrier thickness are required to better understand these effects, our observation of a large TMR in the Co/BFO/LSMO system suggests that the spin disorder present at the BFO/LSMO interface is weak.

X-rays absorption spectroscopy (XAS) measurements have been performed on a BFO(70 nm)/LSMO sample with a photoemission microscope (PEEM) at SIM beamline at the Swiss Light Source. The absorption spectra were measured in electron-yield mode at a grazing photon beam incidence, as shown in fig. 3 (inset). We can exclude ferromagnetic order in our BFO thin films as no circular dichroism has been observed (within experimental accuracy) when recording two XAS with opposite elliptic polarization (see also [22]). Fig. 3 displays two XAS measurements at the Fe $L_{3,2}$ edges obtained with E parallel and perpendicular to the sample plane. The presented absorption spectra have been averaged over an area of 5 µm in diameter. The XAS features are in good agreement with hematite ($\alpha$-$Fe_2O_3$) spectra obtained by P. Kuiper et al. [23] and are similar to the XAS measured for $LaFeO_3$, another AF Fe-based perovskite [24].

While the magnetic origin of circular dichroism in a ferromagnet can be easily demonstrated, the magnetic attribution of linear dichroism is more difficult as it can originate from either anisotropic charge distribution, crystal field or from antiferromagnetism [23]. However, from the similitude of the measured XAS features with the results obtained on $\alpha$-$Fe_2O_3$ with E applied parallel or perpendicular to the AF axis [23], we conclude that the linear dichroism we observe here contains a strong contribution from the AF character of BFO. We also note that Zhao et al have recently shown that in BFO XMLD arises from both ferroelectricity and antiferromagnetism [25]. Finally, we point out that the AF character of our BFO films has been recently confirmed by preliminary neutron diffraction experiments [26].

To exploit the antiferromagnetic behaviour of our BFO layers at room temperature, we have searched for an exchange-bias (EB) effect between a BFO layer and a high $T_C$ ferromagnet (FM) [27]. As the $T_N$ of BFO is high (640K) and because heating a FM/BFO bilayer above this temperature could promote interdiffusion, we did not field-cool the sample through $T_N$. Instead, following Dho et al [28], we have induced a uniaxial anisotropy in the FM by sputtering a 5 nm CoFeB layer on a BFO film at room temperature in a magnetic field of about $H_{growth}$=200 Oe applied in the [100] direction of BFO and a 10 nm capping layer of Au.

Figure 4a shows the M(H) cycle of a test sample of CoFeB(5 nm) grown in the same conditions, on a $SiO_2$/Si substrate. As expected the CoFeB has a square and symmetric hysteresis loop with a low coercive field of ~1 Oe. Figure 4b and c shows the M(H) loop measured at 300K with H either parallel or antiparallel to $H_{growth}$ in the presence of a 35 nm BFO underlayer : the loop is enlarged and shifted along the field axis compared to that of Fig 4a, reflecting the presence of an exchange bias at the CoFeB/BFO interface ($H_c$ = 42 Oe, $H_{eb}$ = -62 Oe). When H is applied perpendicular to $H_{growth}$ (figure 4d.), the M(H) is typical of a measurement along a hard axis, indicating that an anisotropy has been created in the CoFeB layer. We note that we have observed this effect in several samples with different BFO and CoFeB thicknesses, and with other soft ferromagnets such as $Ni_{80}Fe_{20}$. Remarkably, the variation of $H_{eb}$ as a function of the thickness of the BFO pinning layer is in good agreement with what was found for the $Ni_{80}Fe_{20}$/FeMn system [27]. This further confirms that BFO is responsible for the onset of exchange bias in our CoFeB/BFO samples.

We note that exchange bias has already been observed at low temperature with other AFMs, like in $YMnO_3$/$SrRuO_3$ bilayers [29] or $Cr_2O_3$/CoPt structures [30]. Indications of room temperature exchange bias have also been reported on NiFe/Cu/NiFe/BFO spin-valve structures [28]. However, our results represent the first direct observation of exchange bias with a multiferroic material at room temperature. Given the high $T_N$ of BFO (640K), exchange bias should survive up to rather high temperatures. Furthermore, we have verified that the exchange bias effect is stable upon cycling the sample in a magnetic field and shows no training effect [27] (see figure 4e). These are crucial points for the possible exploitation of this exchange bias effect in spintronics devices.

In summary, we have shown that very thin BFO films can be used as tunnel barriers between electrodes of LSMO and Co. This leads to a large positive TMR at low temperature. In addition, we have used the AF character of BFO to induce a robust EB on a CoFeB film at 300K. The next step would now be to reverse the sign of the exchange bias by applying an electric field at room temperature, via the magnetoelectric coupling existing in BFO [25]. This would provide an electrical control of the magnetization of the CoFeB, and thus of the resistance of an MTJ using BFO as the pinning layer.

This work has been supported by the E.U. STREP MACOMUFI (033221) and the contract FEMMES of the Agence Nationale pour la Recherche. H.B. acknowledges financial support from the Conseil Général de l'Essonne. We would also like to thank Andrea Locatelli, Mark Blamire and Ron Jansen for useful discussions.

Figure Captions :

Fig 1. TEM cross section of a BFO(8nm)/LSMO(15nm)//STO (001) sample.

Fig 2. (a) R(H) curve of a Au/CoO/Co/BFO/LSMO//STO (001) 30x30µm² junction measured at 3K with a bias of 10mV. The arrows show the magnetic states of each layer for the different resistance states. (b) Evolution of the TMR with the temperature. The TMR is normalized to its 3K value.

Fig 3. XAS spectra measured in BFO(70nm)/LSMO(15nm)//STO (001) with the linear polarization vector E oriented parallel and perpendicular to the sample plane.

Fig 4. (a) Hysteresis loop along the [100] direction of CoFeB(5nm)//Si layer. Hysteresis loops along (b) the [100] direction, (c) the [-100] direction and (d) the [010] direction of CoFeB(5nm)/BFO(35nm)//STO bilayer. Both structures are capped with a 10 nm Au layer. (e) Coercive fields of a similar sample for successive H cycles (up to ±300 Oe). The exchange field is constant (49 Oe) within the AGFM resolution (~1 Oe). All measurements were made at 300K. Dashed lines are guides to the eye.

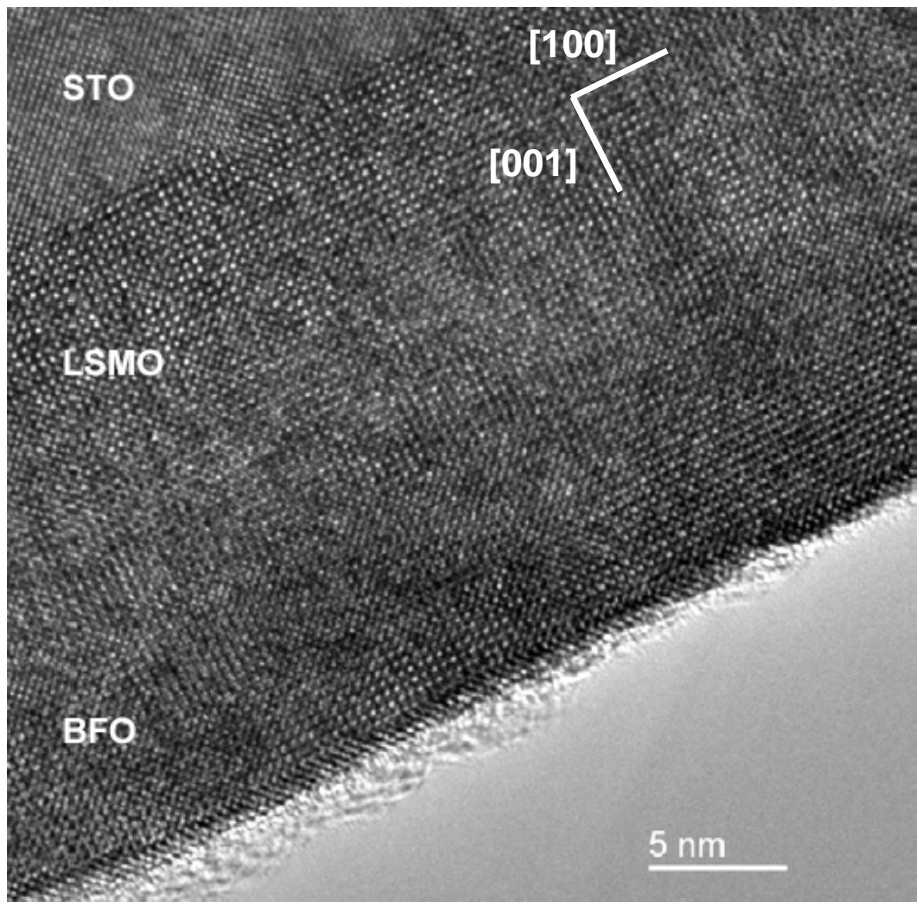

H. Béa et al, Fig. 1

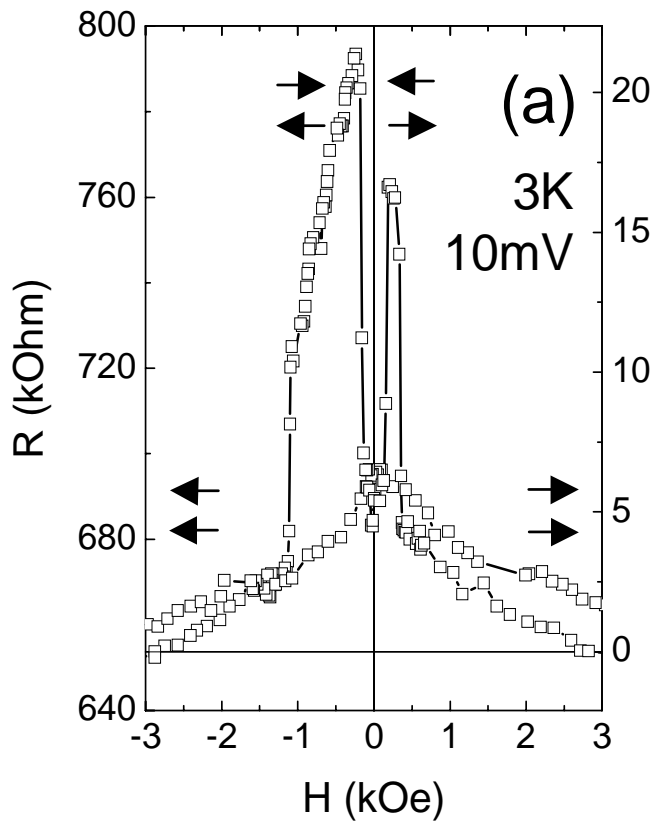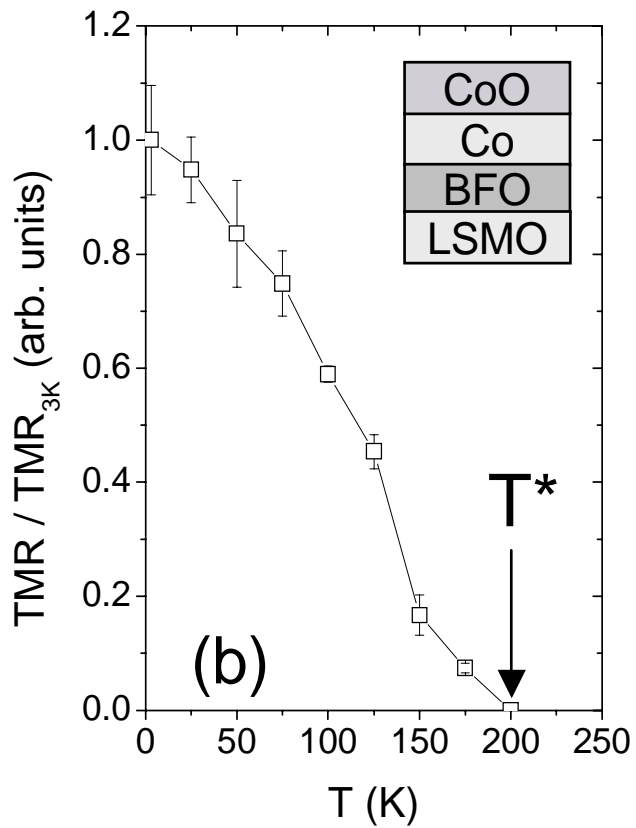

H. Béa et al, Fig. 2

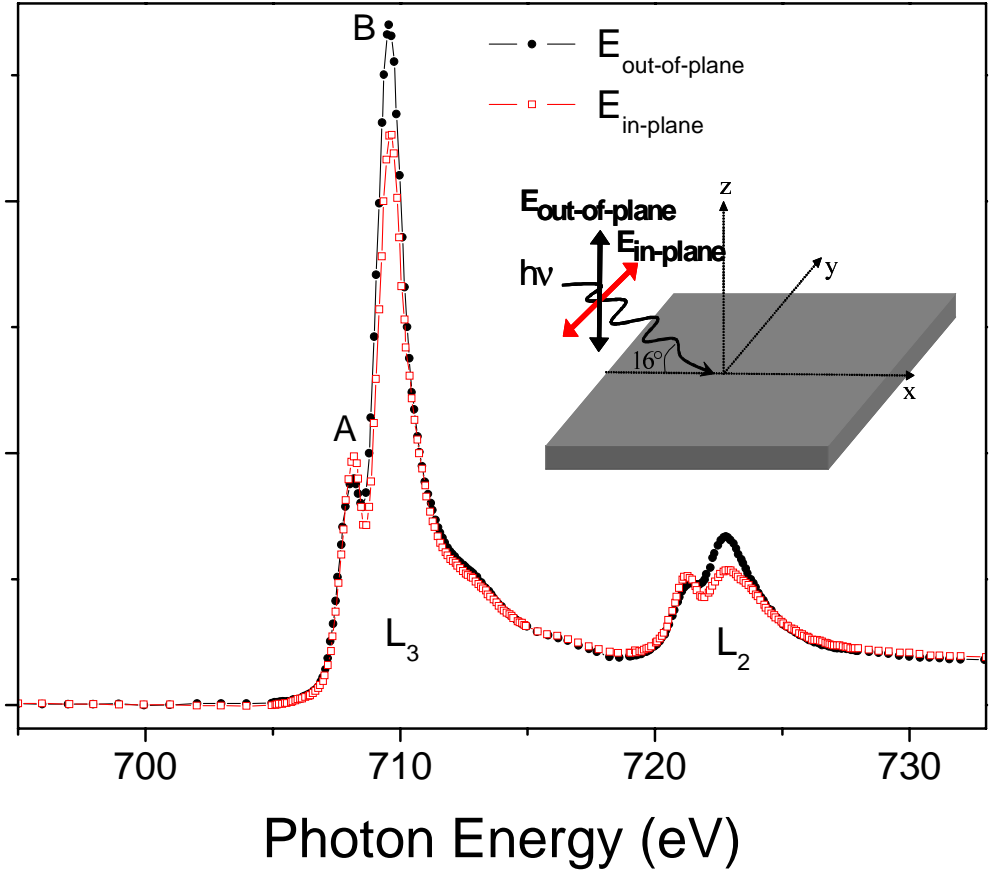

H. Béa et al, Fig. 3

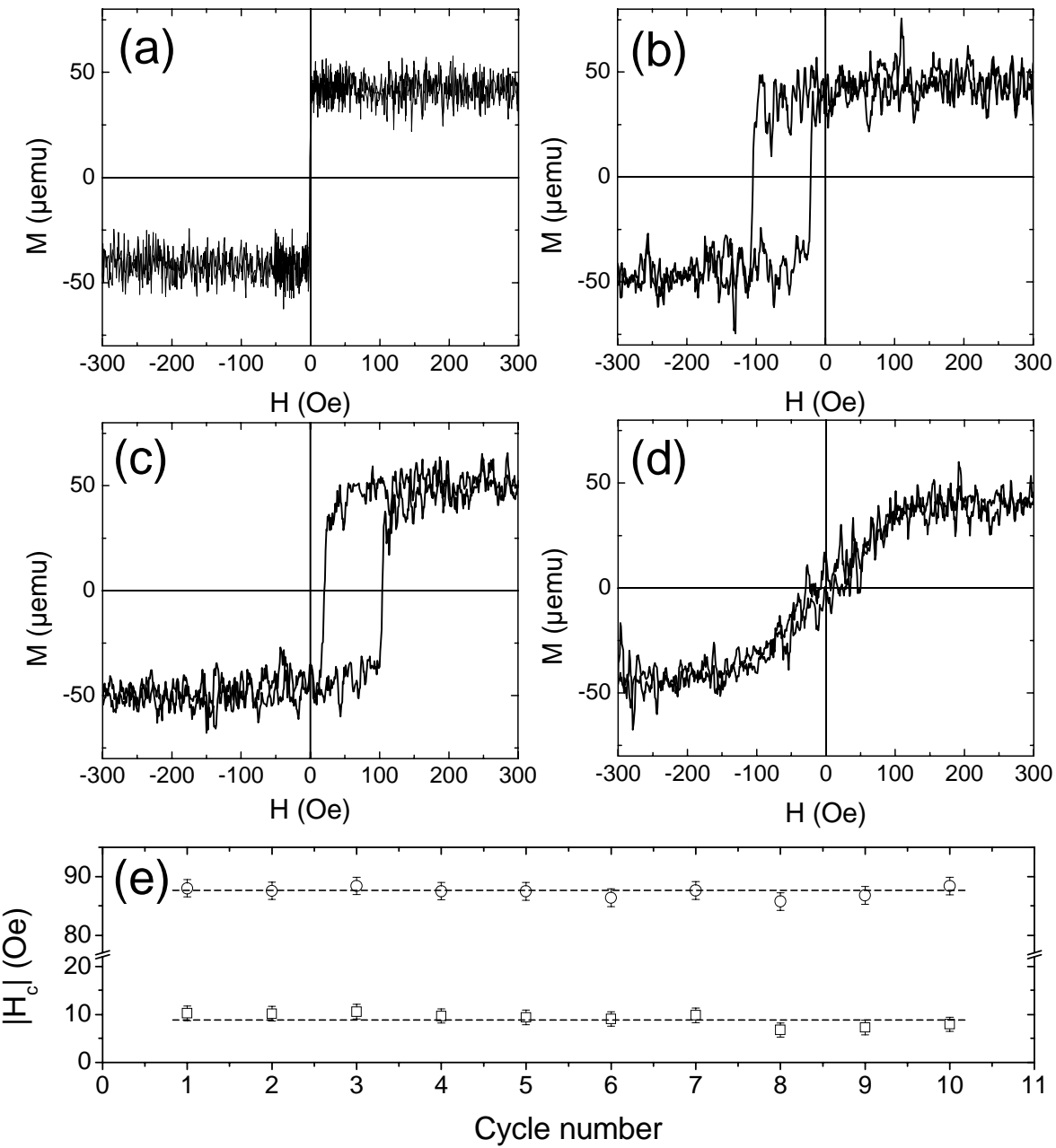

H. Béa et al, Fig. 4